\documentclass{wscpaperproc}
\usepackage{latexsym}
\usepackage{graphicx}
\usepackage{mathptmx}

\usepackage{amsmath}
\usepackage{amsfonts}
\usepackage{amssymb}
\usepackage{amsbsy}
\usepackage{amsthm}

\usepackage{array}
\usepackage{multirow}
\usepackage{hhline}		
\usepackage{subfigure}
\usepackage{mathptmx}
\usepackage{mathtools}
\usepackage{todonotes}
\usepackage{placeins}

\usepackage[pdftex,colorlinks=true,urlcolor=blue,citecolor=black,anchorcolor=black,linkcolor=black]{hyperref}

\newtheoremstyle{wsc}
{3pt}
{3pt}
{}
{}
{\bf}
{}
{.5em}
{}

\theoremstyle{wsc}

\begin{document}

\pagestyle{fancyplain}

\thispagestyle{plain}
\firstPageHead{}

\chead{\fancyplain{}{\itshape Hillmann, Uhlig, Dreo Rodosek, and Rose}}

\rhead{}
\cfoot{}
\renewcommand{\headrulewidth}{0pt} 

\makeatletter
\let\@internalcite\cite
\def\cite{\def\@citeseppen{-1000}%
    \def\@cite##1##2{(##1\if@tempswa , ##2\fi)}%
    \def\citeauthoryear##1##2##3{##1 ##3}\@internalcite}
\def\citeNP{\def\@citeseppen{-1000}%
    \def\@cite##1##2{##1\if@tempswa , ##2\fi}%
    \def\citeauthoryear##1##2##3{##1 ##3}\@internalcite}
\def\citeN{\def\@citeseppen{-1000}%
    \def\@cite##1##2{##1\if@tempswa, ##2)\else{}\fi}%
    \def\citeauthoryear##1##2##3{##1 (##3)}\@citedata}
\def\citeA{\def\@citeseppen{-1000}%
    \def\@cite##1##2{(##1\if@tempswa , ##2\fi)}%
    \def\citeauthoryear##1##2##3{##1}\@internalcite}
\def\citeANP{\def\@citeseppen{-1000}%
    \def\@cite##1##2{##1\if@tempswa , ##2\fi}%
    \def\citeauthoryear##1##2##3{##1}\@internalcite}
\def\shortcite{\def\@citeseppen{-1000}%
    \def\@cite##1##2{(##1\if@tempswa , ##2\fi)}%
    \def\citeauthoryear##1##2##3{##2 ##3}\@internalcite}
\def\shortciteNP{\def\@citeseppen{-1000}%
    \def\@cite##1##2{##1\if@tempswa , ##2\fi}%
    \def\citeauthoryear##1##2##3{##2 ##3}\@internalcite}
\def\shortciteN{\def\@citeseppen{-1000}%
    \def\@cite##1##2{##1\if@tempswa, ##2\else{}\fi}%
    \def\citeauthoryear##1##2##3{##2 (##3)}\@citedata}
\def\shortciteA{\def\@citeseppen{-1000}%
    \def\@cite##1##2{(##1\if@tempswa , ##2\fi)}%
    \def\citeauthoryear##1##2##3{##2}\@internalcite}
\def\shortciteANP{\def\@citeseppen{-1000}%
    \def\@cite##1##2{##1\if@tempswa , ##2\fi}%
    \def\citeauthoryear##1##2##3{##2}\@internalcite}
\def\citeyear{\def\@citeseppen{-1000}%
    \def\@cite##1##2{(##1\if@tempswa , ##2\fi)}%
    \def\citeauthoryear##1##2##3{##3}\@citedata}
\def\citeyearNP{\def\@citeseppen{-1000}%
    \def\@cite##1##2{##1\if@tempswa , ##2\fi}%
    \def\citeauthoryear##1##2##3{##3}\@citedata}
%
%
%
\def\@citedata{%
    \@ifnextchar [{\@tempswatrue\@citedatax}%
                  {\@tempswafalse\@citedatax[]}%
}

\def\@citedatax[#1]#2{%
\if@filesw\immediate\write\@auxout{\string\citation{#2}}\fi%
  \def\@citea{}\@cite{\@for\@citeb:=#2\do%
    {\@citea\def\@citea{, }\@ifundefined
       {b@\@citeb}{{\bf ?}%
       \@warning{Citation `\@citeb' on page \thepage \space undefined}}%
{\csname b@\@citeb\endcsname}}}{#1}}%

%
\def\@citex[#1]#2{%
\if@filesw\immediate\write\@auxout{\string\citation{#2}}\fi%
  \def\@citea{}\@cite{\@for\@citeb:=#2\do%
    {\@citea\def\@citea{, }\@ifundefined
       {b@\@citeb}{{\bf ?}%
       \@warning{Citation `\@citeb' on page \thepage \space undefined}}%
{\csname b@\@citeb\endcsname}}}{#1}}%

%
\def\@biblabel#1{}
\makeatother



\newdimen\bibindent
\bibindent=0.0em
\def\thebibliography#1{\section*{\refname}\list
   {}{\settowidth\labelwidth{[#1]}
   \leftmargin\parindent
   \itemindent -\parindent
   \listparindent \itemindent
   \itemsep 0pt
   \parsep 0pt}
   \def\newblock{}
   \sloppy
   \sfcode`\.=1000\relax}


\setlength{\baselineskip}{12.7pt}

\title{SIMULATION AND OPTIMIZATION OF CONTENT DELIVERY NETWORKS CONSIDERING USER PROFILES AND PREFERENCES OF INTERNET SERVICE PROVIDERS}
\author{
Peter Hillmann\\
Tobias Uhlig\\
Gabi Dreo Rodosek\\
Oliver Rose\\[12pt]
Department of Computer Science\\
Universit\"at der Bundeswehr M\"unchen\\
Werner-Heisenberg-Weg 39\\
Neubiberg, 85577, GERMANY\\
}

\maketitle

\section*{ABSTRACT}
A Content Delivery Network (CDN) is a dynamic and complex service system. It causes a huge amount of traffic on the network infrastructure of Internet Service Providers (ISPs).
Oftentimes, CDN providers and ISPs struggle to find an efficient and appropriate way to cooperate for mutual benefits.
This challenge is key to push the quality of service (QoS) for the end-user. 
We model, simulate, and optimize the behavior of a CDN to provide cooperative solutions and to improve the QoS.
Therefor, we determine reasonable server locations, balance the amount of servers and improve the user assignments to the servers. 
These aspects influence run time effects like caching at the server, response time and network load at specific links. Especially, user request history and profiles are considered to improve the overall performance.
Since we consider multiple objectives, we aim to provide a diverse set of pareto optimal solutions using simulation based optimization. 

\section{INTRODUCTION}\label{introduction}
Content Delivery Networks (CDN) are large, distributed systems of servers that provide a storage infrastructure.
They represent a network platform to provide data services to end-users with high availability and high performance. Providers of a CDN offer these infrastructures to content providers with intention to deliver their information to end-users.

The servers of the CDN copy and cache the content of the original source, to improve the network and server performance. Effectively, a CDN reduces the need for content providers to own and mange a dedicated distribution infrastructure for their content.
We focus on collaborative and community driven CDNs that replicate data based on demand independently of its source.
CDNs process large number of requests leading to a high amount of transmission data transferred by the Internet Service Provider (ISP). To avoid bottlenecks, severs should be placed at favorable locations with appropriate network connection. Especially, positioning servers at gateways of providers of other network infrastructures ensures the scalability and reliability of the system. At the same time, the length of the network path from an offering source to the requesting destination should be minimized. Finally, the strategic assignment of requests to nearby and appropriate server can significantly improve the load on the server and the network. From an end-user perspective, we obtain a higher performance and reliable quality.

A collaboration between CDN and ISP is necessary to enable an efficient network platform as a service for content providers and users. The challenge is to set up a cooperation that mutually benefits both parties. This paper introduces a concept for an effective CDN-ISP collaboration. It supports the planning of a CDN including its server locations and storage properties as well as the user assignment at runtime. Our approach is evaluated by simulating the basic processes of a CDN using a given ISP infrastructure. We demonstrate that a cooperation lead to improvements for both parties. Furthermore, we evaluate the behaviors of the users according request profiles based on their request history.

This paper is structured as follows: In Section 2 we describe a typical scenario and the requirements for a CDN-ISP collaboration. Section 3 provide an overview of other approaches in that application area. The main part in Section 4 describes our model of the CDN and the simulation-based optimization approach. Subsequently, we explain the process of optimization, evaluate our concept, and identify fundamental properties of the presented system in Section 5 and 6. The last section summarizes our work and provides and outlook.

\section{SCENARIO AND REQUIREMENTS}\label{scenario}
\begin{figure}[hbtp]
{
\centering
\includegraphics[width=1.0 \textwidth]{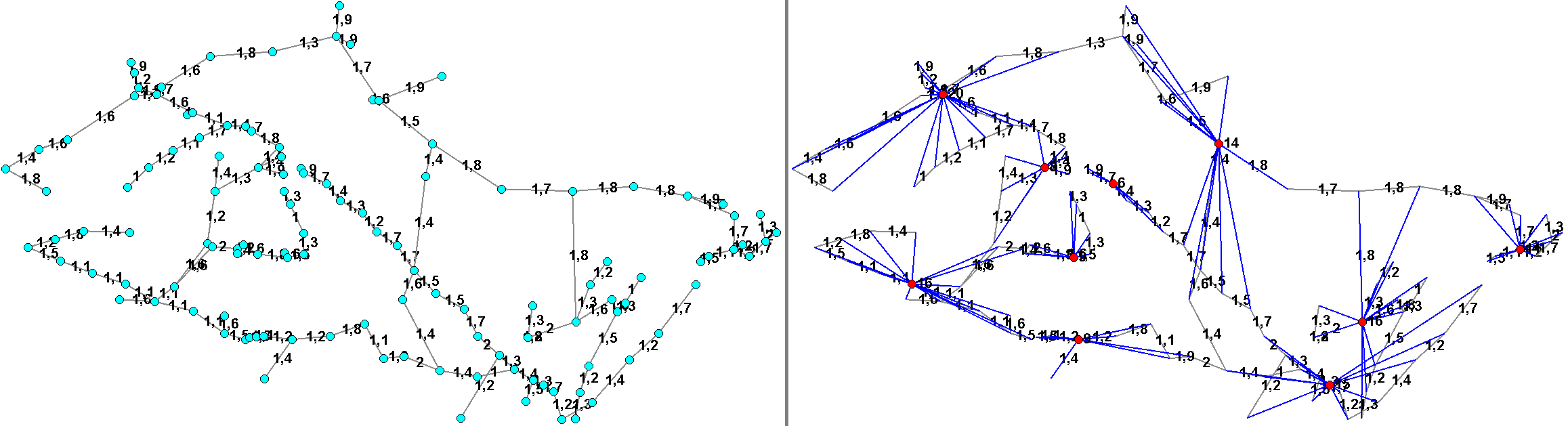}
\caption{Example of the abstract model with a non-hierarchical network infrastructure including 126 weighted edges and 124 grouped users to 10 prioritized network nodes. In this case, the network nodes are simply assigned to the closest server.}
\label{fig:AbstractScenario2}
}
\end{figure}

The management of a CDN intends to improve their infrastructure and service. In consultation with the responsible ISP, a predefined amount of mirror servers has to
be placed and connected strategically to the ISP network infrastructure. An example setup is shown in Figure \ref{fig:AbstractScenario2}, the network of  Dialtelecom \shortcite{Knight2011}. The left side presents the geographical location of the network nodes and the edges of the infrastructure. The right side includes the registered mirror servers. The blue lines represent the assignments of the network nodes to their closest server, i.e., the intended user allocation.\\

However, the closest server may not be the best one to satisfy a given user request, depending on the available data at the server, the load on the server and the current network performance.
With consideration to multiple relevant objectives, the following questions and criteria describe the most common management challenges from CDN providers and ISPs: 
\begin{itemize}
\item Server-Placement: Where should servers be placed in the network infrastructure? Usually, servers should be placed close to the end-users to improve QoS \textit{(CDN)} and reduce network load as well as lower the risk of bottlenecks \textit{(ISP)}.
\item User-Assignment: Which user is assigned to which server? End-users should be assigned to a nearby server with suitable content to get a cache hit and quick response \textit{(CDN)}. This avoids redirecting and minimizes network load caused by data access from another source\textit{(ISP)}. 
Generally, we may accept larger distances to the servers if we can expect higher cache hit probabilities. 
\item Server-Amount: What is the necessary amount of servers? An appropriate number of server enables a specific Service-Level with minimal cost of operation \textit{(CDN)} and reduces transition costs \textit{(ISP)}. 
\item Cache-Size: How much storage do we need for a certain server to obtain an effective system? A suitable amount of caching space improves the storage behavior and balances the operational costs \textit{(CDN)}. It also lowers the amount of redirects to other sources \textit{(ISP)}.
\end{itemize}

\begin{figure}[hbtp]
{
\centering
\includegraphics[width=0.26 \textwidth]{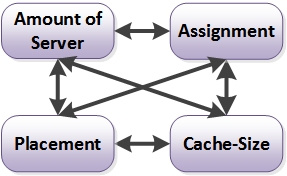}
\caption{Connection of the different optimization criteria in context of the application area CDN.}
\label{fig:OptimierungsProblem}
}
\end{figure}

Figure \ref{fig:OptimierungsProblem} illustrates the dependencies between these aspects. All of them directly or indirectly impact the other ones. For example, the server-amount influences directly the server-placement and the user-assignment. Furthermore, the cache-size influenced indirectly. 

\textit{Circle of conflict:} A higher amount of servers results in an improved placement with shorter network paths. The average connection distances get reduced and QoS is increased. User have access to more nearby servers and thereby more potential assignments. A single server receives fewer requests.
The explicit need for large cache sizes is reduced, however, this leads to an increased amount of cache misses.
In contrast, a reduced amount of servers will lead to opposite effects.\\

The server-placement and the user-assignment are each considered to be NP-hard problems. To find suitable solutions, we need to optimize them simultaneously, because they are strongly connected to each other.
The position of a server is influenced by the network infrastructure and the assigned user. Every user should have a minimized connection length to his assigned server. 
The group of assigned users to a server depends on their user profiles and the network path. 
The grouped profiles should be positively correlated so that the server obtains requests for the same data.
Therefore, the best server position in relation to a short network connection does not necessary result in a group assignment with positive correlated profiles.
Furthermore, the adequate amount of server has to be analyzed.
Equation \ref{form:distance} and Equation \ref{form:profil} are the objective function for the complex optimization problem, where $user_i$ is assigned to $server_u(user_i)$.
To provide a practical tool for CDNs and ISPs, we must determine a whole pareto front of possible solutions to identify the sweet spot for an effective cooperation.\\

\begin{equation}
Distance = \underset{ } { min }\text{ } \underset{user_i = 1, ..., n }{ max } \text{ } distance(user_i|server_u(user_i))
\label{form:distance}
\end{equation}
\begin{equation}
Correlation = \underset{ } { max }\text{ } \sum_{  user_i = 1 }^n { } \text{ } \rho(user_i|server_u(user_i))\\
\label{form:profil}
\end{equation}

\section{RELATED WORK}\label{relatedwork}
Previously, some solutions for an effective CDN-ISP cooperation have been proposed. 
The work of \citeN{Pallis:2006:IPC:1107458.1107462} presents an overview on CDNs and identifies the most common practices. 
It conveys a rough idea on balancing costs of CDN providers and content providers while respecting the QoS for Web customers. However, it does not consider the role of ISPs and lacks any detailed suggestions for reaching typical optimization objectives. Nevertheless, their work highlights the importance of appropriate server placement and content management for servers. They, especially, identified dynamic caching-related processes for CDNs as an area with great optimization potential. 
\citeN{Mobasher:2000:APB:345124.345169} cluster users according to their profile and the history of URL demands. These clusters are used to predict future HTTP requests. However, their work considers only a single Web Server of a company. We extend the idea of user profiles to a large scale system for a CDN-ISP cooperation to improve the overall performance.
The publication of \citeN{Qiu2001} addresses placement strategies for Web Server replicas in contrast to redirecting requests. They use workload information to save appropriate content on already existing mirror servers to improve access latency and load balance. They also discuss practical issues like imperfect information. Nevertheless, this work is limited to the network aspect of a complex system without taking advantage of user profiles.
One of the most influential paper is from \shortciteN{Frank:2013:PCC:2500098.2500103}.
Their concept allows a CDN to incorporate recommendations from an ISP to assign an end-user to a server and to determine locations of servers and content as well as the required amount of servers. The advantage of their system is that no routing changes are necessary to reduce operational costs. Their main focus is on the communication and management processes between CDN Provider and ISP, but not on solving the underlying challenges. They also ignore dynamic problem aspects since they do not use an analytical simulation model.
The ALTO protocol \shortcite{R.Alimi2014} is an application layer protocol according to the ISO/OSI seven-layer model. It provides information that are usually hidden, including: network topologies, link availability, routing policies, and path costs from a cooperating ISPs. This information is required to implement and realize our approach in practice.

The work of \shortciteN{Jiang:2009:CCD:1555349.1555377} demonstrates that considering server selection and traffic engineering separately leads to sub-optimal solutions. It serves as a starting point to better understand these interaction for those that operate networks and those that distribute content. Their static system does not take user profiles into account. They purely focus on the network costs.
\shortciteN{Krishnan:2009:MBE:1644893.1644917} investigated latency-based server selection and redirection to improve the CDN performance. This analysis does not consider the effects of dynamic caching behavior at the server and whether requested content is actually available at the specific server.
Overall, these approaches used a static system description ignoring the dynamic effects in user behavior driven CDN. Furthermore, they focus more on economic and financial interests without considerations of an improved quality of service for the user.
In contrast to the existing work, we analyze the CDN-ISP collaboration with dynamic effects using simulation-based optimization. Therefore we are able to consider run time effects and optimize according to multiple objectives.

\section{OPTIMIZATION AND SIMULATION MODEL}\label{Approach}
Our generic model is based on an existing network infrastructure. We use publicly available data from the \textit{Internet Topology Zoo} \shortcite{Knight2011}. Nevertheless, the approach is adaptable to any other graph structure. For an effective and reliable service, the mirror servers of the CDN should be connected strategically to the network infrastructure. A server should be placed close to the end-user to provide a continuous service with respect to the network path. This also avoids bottlenecks and queuing delays during transmissions. Different links are weighted according to their properties like bandwidth, latency and utilization. The necessary information is provided by the ISP. Due to the expected high amount of served requests from distributed end-users, the servers need a high-performance connection. This is provided by the backbone networks in the TIER 1 or 2 topology. Therefore, we focus on Top-Level infrastructures to identify optimized mirror server locations. A server is always connected to an existing network node in the given infrastructure. The short link from a server to the selected node is negligible in relation to a long network path to the end-user.
Additionally, locations can be prioritized depending on other aspects like operating costs and performance. The resulting placement constraint corresponds to the placement of a mirror server at a weighted node from the backbone topology respecting connection distances. We consider a placement of a server at a data link or in an area without direct connections as impractical.

Usually, the end-users are connected in the TIER 3 topology, which has a tree like structure. Effectively, the users share a common internet gateway, for example one for a street, or one for a region. The necessary structural information has to be provided by the ISP. It includes the geographical location of the end-user connection to the internet gateway as well as the local infrastructure. To reduce the amount of optimization and simulation objects, end-users are aggregated. Multiple end-users connected to the same access node are grouped to represent the respective node. This level of abstraction is precise enough, because every CDN request from such a group has to pass their common gateway node. All end-user properties and behaviors are also combined. The demands of the users are accumulated to a group profile.
To differentiate between groups of end-user, the network node is assigned a priority and a request profile. The priority depends on aspects like the amount of users, their economical relevance, and frequency of requests. The request profile reflects the history and prevalence of requested data and URLs. The required information is usually tracked by the CDN provider to predict future requests and to improve service \cite{Mobasher:2000:APB:345124.345169}.
We model the end-user profiles using recorded, request traces from file access and cache benchmarks \cite{headissue2000}.
Alternatively, we generate pseudo realistic profiles using the Zipf-distribution \shortcite{Breslau1999}. Each modeled profile entry consists of the name of service and the probability that the user requests it at a specific time. A profile can have multiple entries.

\begin{figure}[hbtp]
{
\centering
\includegraphics[width=0.6 \textwidth]{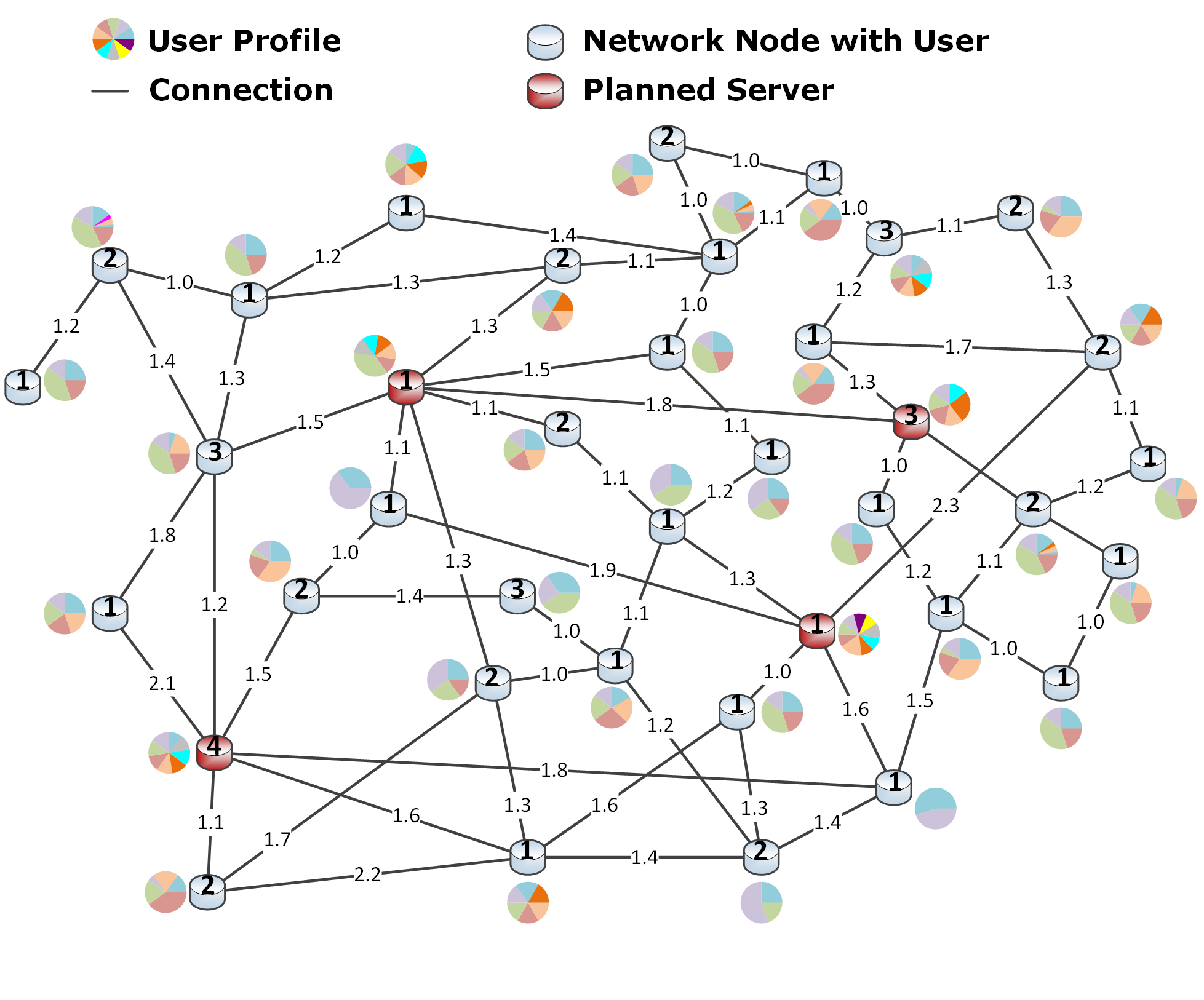}
\caption{Example of the abstract model with a non-hierarchical network infrastructure including weighted edges and grouped users to a prioritized network node.}
\label{fig:AbstractScenario}
}
\end{figure}

An example scenario is visualized in Figure \ref{fig:AbstractScenario}.
It includes prioritized network nodes for groups of users and for placement opportunities of servers. The network nodes are connected with each other via several, weighted links using the existing infrastructure. The requested content is described by the profile of the network node and represented by a pie chart.\\

During optimization, we place a predefined amount of servers in the network and assign the user groups to their appropriate server. The server placement incorporates multiple aspects. The shortest network path between server and user are calculated with Dijkstra's algorithm. We also factor in priorities, weights and profile correlation. Accordingly, we consider the following optimization objectives: Minimization of maximum or average distance between user and assigned server and profile correlation based on Spearman's rank correlation coefficient.

In our simulation, the end-users send requests to their server. The server is responsible to fully answer the request with the desired data. This specific behavior is dependent on the organization and configuration of the CDN. More complex types with redirecting requests will be considered in future. To answer a request, the necessary data have to be available on that specific server. Otherwise, it is fetched  directly from the source or another server in the CDN. In addition to answering the user requests, a server manages its cache and the provision of missing data. To simulate this behavior, we model a cache at servers with a predefined size and cache replacement strategy, e.g., LRU-$X$, LFU, LRFU \shortcite{Lee:2001:LSP:626527.627193}, and LIRS \cite{Jiang:2002:LEL:511334.511340}. The cache stores the readily available data and tracks statistical values like miss rate. The optimal caching behavior is calculated according to \textit{Belady} \cite{Belady:1966:SRA:1663374.1663376} afterwards.
For simplification purpose, we assume uniform data size.
During Simulation we track the network load and the connection distance of the communicating entities. A high QoS is reached by combining a high cache hit ratio and a short transmission path --- our main objectives.

\begin{figure}[hbtp]
{
\centering
\includegraphics[width=1.0 \textwidth]{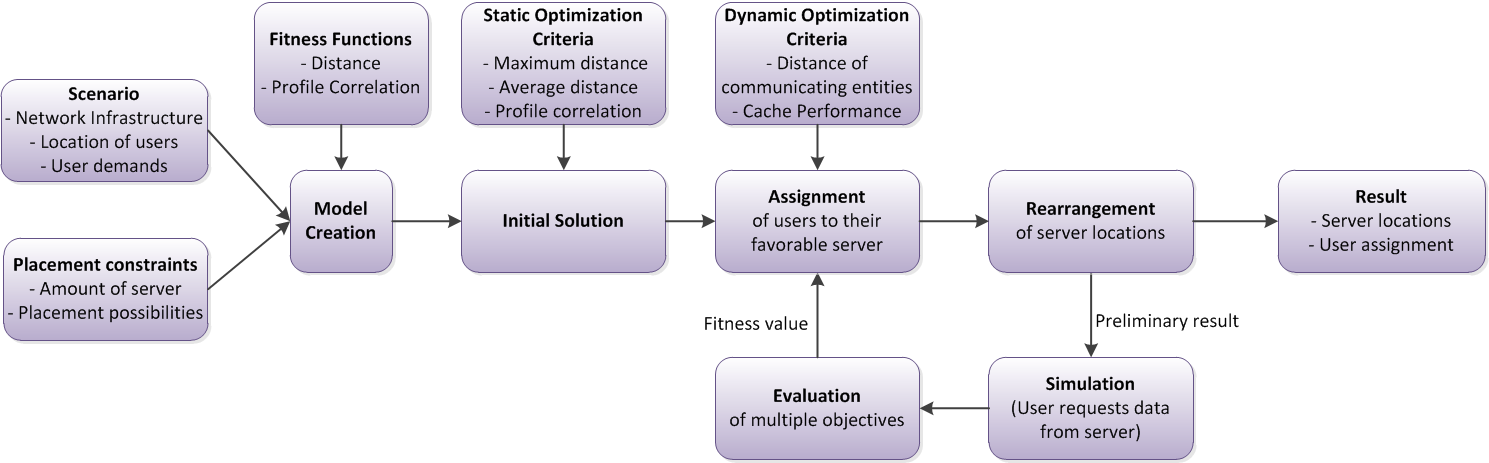}
\caption{Processing scheme to optimize the placement of mirror server.}
\label{fig:Modelling}
}
\end{figure}

The process of model creation, optimization and simulation is presented in Figure \ref{fig:Modelling}.
The data of the scenario is combined with the placement constraints and enriched with the priorities and weightings of CDN provider and ISP. The created model contains the network infrastructure, locations of users and their demanding profile. Furthermore, it includes information on the specified amount of servers and the specific objectives. During the placement optimization, servers are added to the model and rearranged to improve the performance.
During initialization, we consider only static criteria, which can be obtained without simulation.
Afterwards, the model is simulated and evaluated. In an iterative process, the server locations and the user assignments are optimized. This is done with respect to the aforementioned static and dynamic optimization objectives.
The result includes amongst other things the server locations and the assignment of users to their preferable server.\\

\section{OPTIMIZATION}
For the initial solution, we focus on an optimized server placement without paying attention to the user profiles. It represents the ideal case with a high locality of similar user profiles. Afterwards, the solution is modified stepwise with regard to fitting profiles. The server locations have to be identified with respect to the infrastructure. In intention to equal important users, we minimize the maximum distance from a server to its assigned user. We use the placement algorithms presented in the next section and additionally use an AMOSA algorithm \shortcite{SanghamitraBandyopadhyay2008} as reference.

\subsection{Placement Algorithm}
We designed an algorithm to find optimized locations for server in a given topology, called Dragoon (Diversification Rectifies Advanced Greedy Overdetermined Optimization N-Dimensions). It uses of two stages: initialization and iterative improvement.

As first step of initialization, an orientation mark is placed at the optimal center position 
assuming a single center placement problem - calculate all possible constellations for one server. 
The orientation mark is only used for the placement decision of the first server and discarded afterwards.
The first server is placed at the position of a network node which is farthest away from the orientation mark.
Subsequently, the remaining number of the pre-defined amount of server is placed using the 2-Approx strategy.
It calculates for every network node the distance to all placed server.
Afterwards, it chooses the network node with the largest distance to its closest server as the next location to place an additional server.
Thereby, we obtain a specific solution of the 2-Approx placement strategy, which originally places the first node randomly.

After the initialization, the algorithm starts with the iterative refinement rule to further improve the server locations.
For every server, it tests all connected network nodes from the considered infrastructure with a direct edge to the current position of a server.
If the new location improves the overall situation, the algorithm shifts the server to the better position.
This is done with respect to the specified optimization objectives.

In each iteration step, all network nodes of the observed infrastructure are (re)assigned to their closest server. Every
server is allowed to shift its position only once in each iteration. This iterative optimization is repeated until all server do not change their positions any more. Since the algorithm
accepts only improved positions in every step, it is guaranteed to terminate ultimately.
These optimized locations for server represent central nodes in a given network topology. Other clustering algorithms like MacQueen \cite{MacQueen67}, Lloyd \cite{Lloyd82}, Greedy \cite{Jamin01} and Integer Linear Programming \cite{Kleinberg2013} were evaluated and used as references.

\subsection{Profile Correlation Algorithm}
The spectrum of the user interests is very large. If every service generate a dimension in the multi modal search space, the distance between two profiles become meaningless. So we use a different approach to obtain the Pareto Front starting from optimized placement.

We use an adapted Greedy algorithm to optimize the initial solution towards a profile correlation. It improves the assignment of the users to its preferable server. Therefore, the algorithm focus on the improvement of increased correlation coefficients at every server. The server location is adjusted afterwards. All user profiles assigned to a server are combined to a server profile. A possible reassignment of a user can have a large impact on the correlation. The coefficient represents the fitness of all user requests arriving the same server. So, we include the profile of the currently considered user into the future server profile on a reassignment. For the evaluation, the Spearman's rank coefficient is used as fitness value. 
The following example explains the calculation of the correlation coefficient. A user profile is comparable with entries in a table, which consists of the name of a service and the probability to be requested. In this example, user 1 has the profile \{A=50\%, B=50\%, C=0\%\} and user 2 has the profile \{A=30\%, B=0\%, C= 70\%\}. Both user are assigned to the same server. The aggregated server profile is \{A=40\%, B=25\%, C=35\%\}. According to the rang correlation, user 1 obtain a coefficient of $\rho$=0.125 and user 2 of $\rho$=0.5.

For every user, the algorithm calculates the correlation coefficient with every server. If the correlation coefficient to the new server is positive, it compares the value with the coefficient to the current server.
The user will be reassigned on an improved correlation coefficient at the end of each iteration. All users are simultaneously reassigned to their improved server with the best correlation coefficient.
This process is repeated until no reassignment take place. This process will always terminate, as we only accept improved constellations.
Other heuristics algorithms like Genetic Algorithms and Simulated Annealing are also adapted to this problem and used for comparison. Furthermore, different correlation coefficients are tested.

\section{SIMULATION AND ASSESSMENT}\label{simulation}

We evaluated our approach based on simulation experiments using scenarios from the Internet Topology Zoo \shortcite{Knight2011}. The detailed results depend strongly on the specific scenario and and its network infrastructure. Although, we tested many scenarios we will focus our discussion on the example scenario presented in Section 2.
Generally we observed comparable results for the other scenarios. With regard to the user profiles, the universe has 100 different services and we used the Zipf's distribution with  $\alpha = 0.3$ for profile generation, a typical value to model web traffic. For a simulation run each user sends at least 100 request to its server. \\

\begin{figure}[hbtp]{
\centering
\includegraphics[width=1.0 \textwidth]{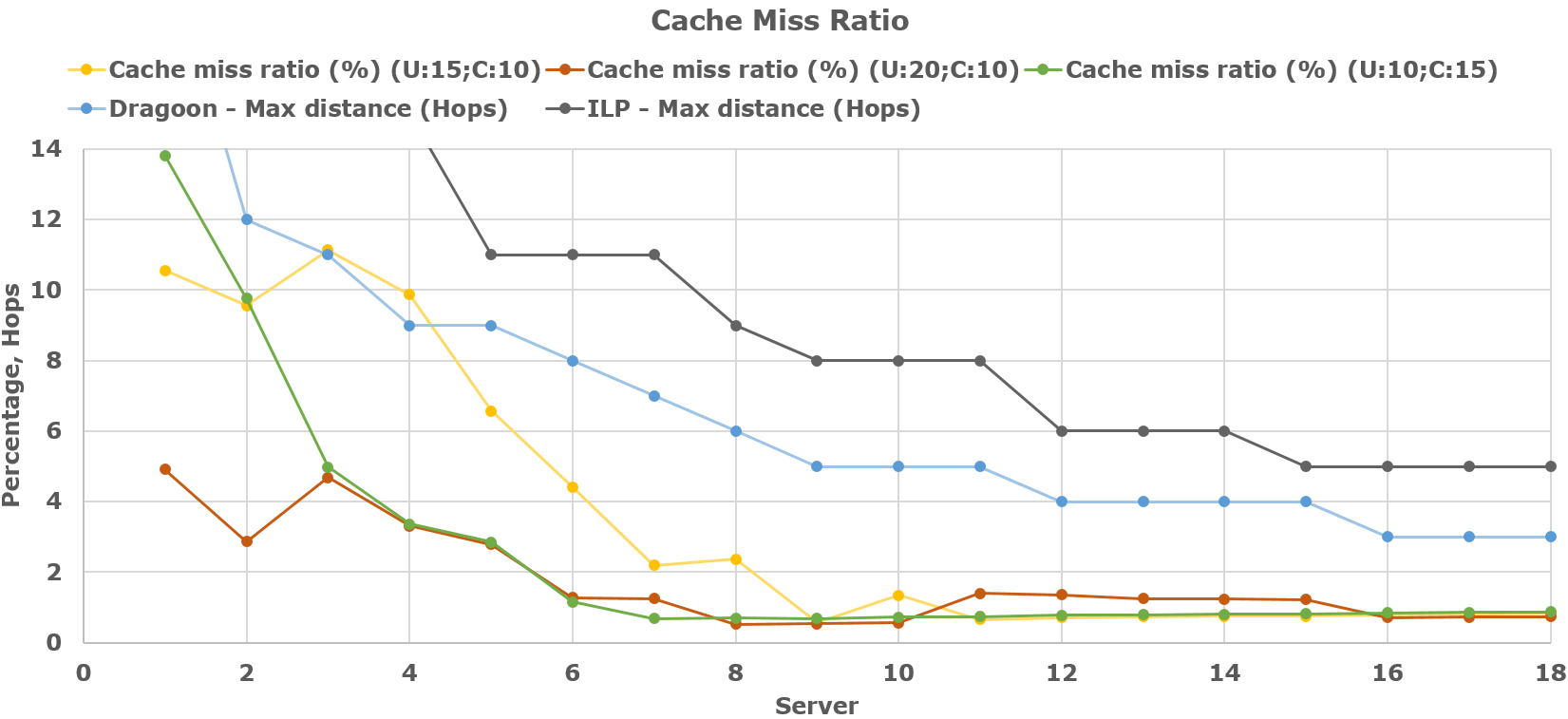}
\caption{Simulation of the cache miss ratio for different amount of server using the example scenario in Section 2. A user profile consists of \textit{U} services and a server buffer up to \textit{C} services.}
\label{fig:exp1}
}\end{figure}

\textbf{Experiment 1:} Initially, we optimized the server placement to a reduced maximum distance from a user to its closest server. During the simulation, every user requests its closest server and has to get the answer from it. Figure \ref{fig:exp1} show the maximum hop distance for different amount of center nodes, blue and gray line.

With growing numbers of centers, the optimization algorithms Dragoon and Integer Linear Programming (ILP) are steadily able to reduce the maximum distance. Dragoon shows a significantly better performance so it is used for the further evaluation. Additionally, the figure compares the cache miss ratio in percent with different simulation parameter. The cache replacement strategy is calculated according to the optimal decision with \textit{Belady}.

Generally, a low amount of servers results in a higher cache miss ratio. Nevertheless, not every additional server improves the caching behavior. An additional server can lead to a negative profile correlation of user requests at the server, which increase the cache miss ratio. This effect is heavily visible for low amounts of server and larger user profiles than server caches.
The slight increase for large amount of servers can be attributed to a higher number of initial cache misses.\\

\begin{figure}[hbtp]{
\centering
\includegraphics[width=1.0 \textwidth]{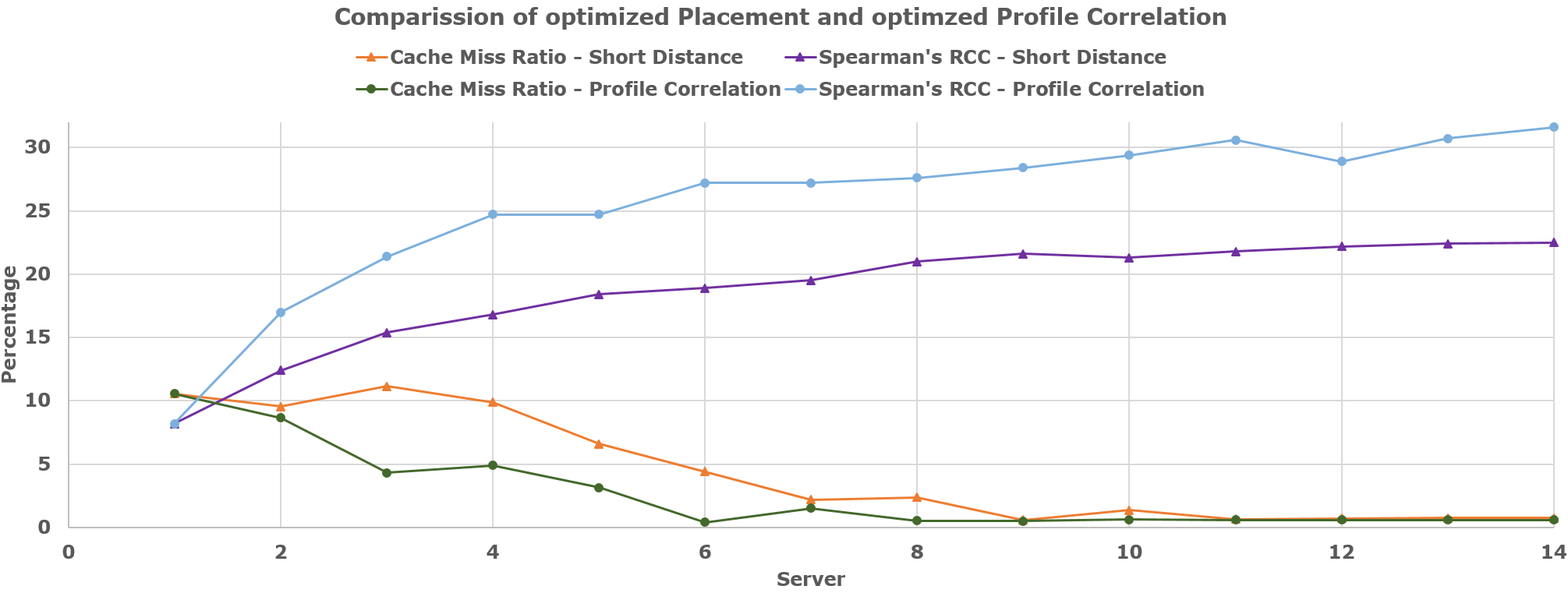}
\caption{Optimization of high profile correlation and simulation of the cache miss ratio for different amount of server using the example scenario in Section 2.}
\label{fig:exp2}
}\end{figure}

\textbf{Experiment 2:} In this simulation, we improve the correlation of the user profiles to the respective server regardless of the network distance.
After the users are clustered into groups and assigned to a server, the specific location of the server is recalculated to minimize the maximum distance to every assigned user. In comparison to Experiment 1, we obtain a local optima for the minimized cache miss ratio for every specified amount of servers. Figure \ref{fig:exp2} shows the connection between profile correlation and cache miss ratio as well as the impact of different amount of servers. 
A server caches up to 10\% and a user requests 15\% of the services in the universe. We illustrate the gap between optimized distance and optimized profile correlation.

The curves show the potential benefits for the CDN provider or ISP depending on the optimization criteria. It highlights the advantages of a cooperation for a common system approach. The optimization towards profile correlation leads to an improved cache hit ratio. With an increased amount of servers, the profile correlation coefficient stagnates or gets worse. This is due to the performance of the profile optimization algorithms and the complexity of the problem. 
Nevertheless, the impact on the cache is minimal due to the already improved caching behavior and high amount of servers. For small amount of servers, the caching behavior is improved heavily. In some situations, the cache miss ratio can be reduced to half and even more. Additionally, an improved profile correlation does not automatically lead to an improved caching behavior in simulation, for example shown in the case from 3 to 4 servers due to cache size.
The average distance from a user to the server is higher and varies more if we optimize according to profile correlation instead of distances.
These results support the provider in decisions for additional servers to improve the system behavior and service level to users.\\

\begin{figure}[hbtp]{
 \centering
 \includegraphics[width=1.0 \textwidth]{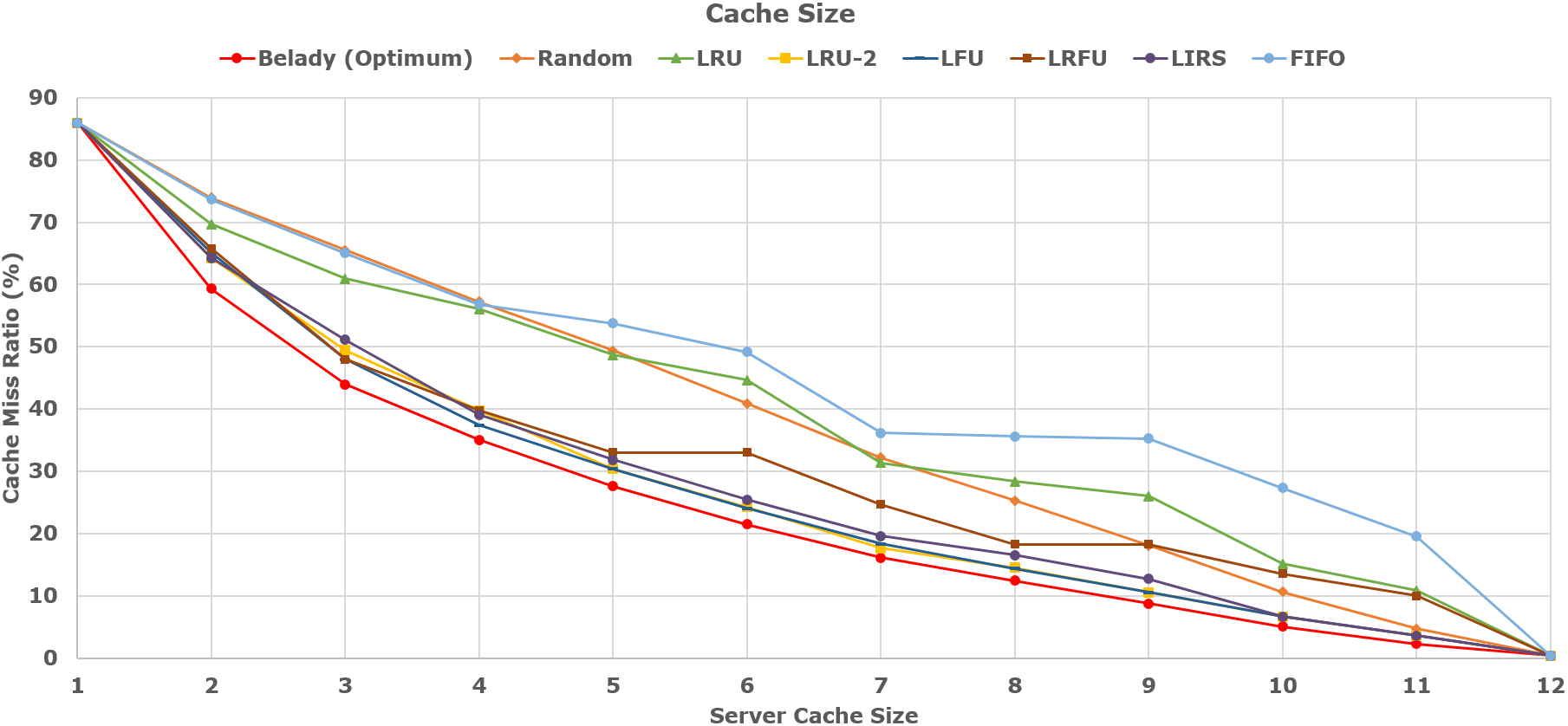}
 \caption{Cache miss ratio for different sizes of cache.}
 \label{fig:exp3}
 }\end{figure}

\textbf{Experiment 3:}
In this experiment, we analyzed the effective size of the cache and evaluate an adjusted replacement algorithm. The scenario uses 5 servers, their location and user assignments are fixed. Every server has the same amount of cache and the user profile contains 15 services. Figure \ref{fig:exp3} shows the cache miss rate for different sizes of cache. It compares the performance of several cache replacement algorithms.

With increased size of cache, the improvement is reduced with every additional cache space until it stagnates. LRU-2, LFU and LIRS are close to the global optimum calculated in accordance with Belady. With a cache size of 12, we reach the minimum, whereas increased cache size has negligible impact on the cache miss ratio. Even if the spectrum of a user profile has more entries, the additional services are not requested very often due to Zipf-Distribution. So it has negligible impact on the general caching behavior.\\

\begin{figure}[hbtp]{
\centering
\includegraphics[width=1.0 \textwidth]{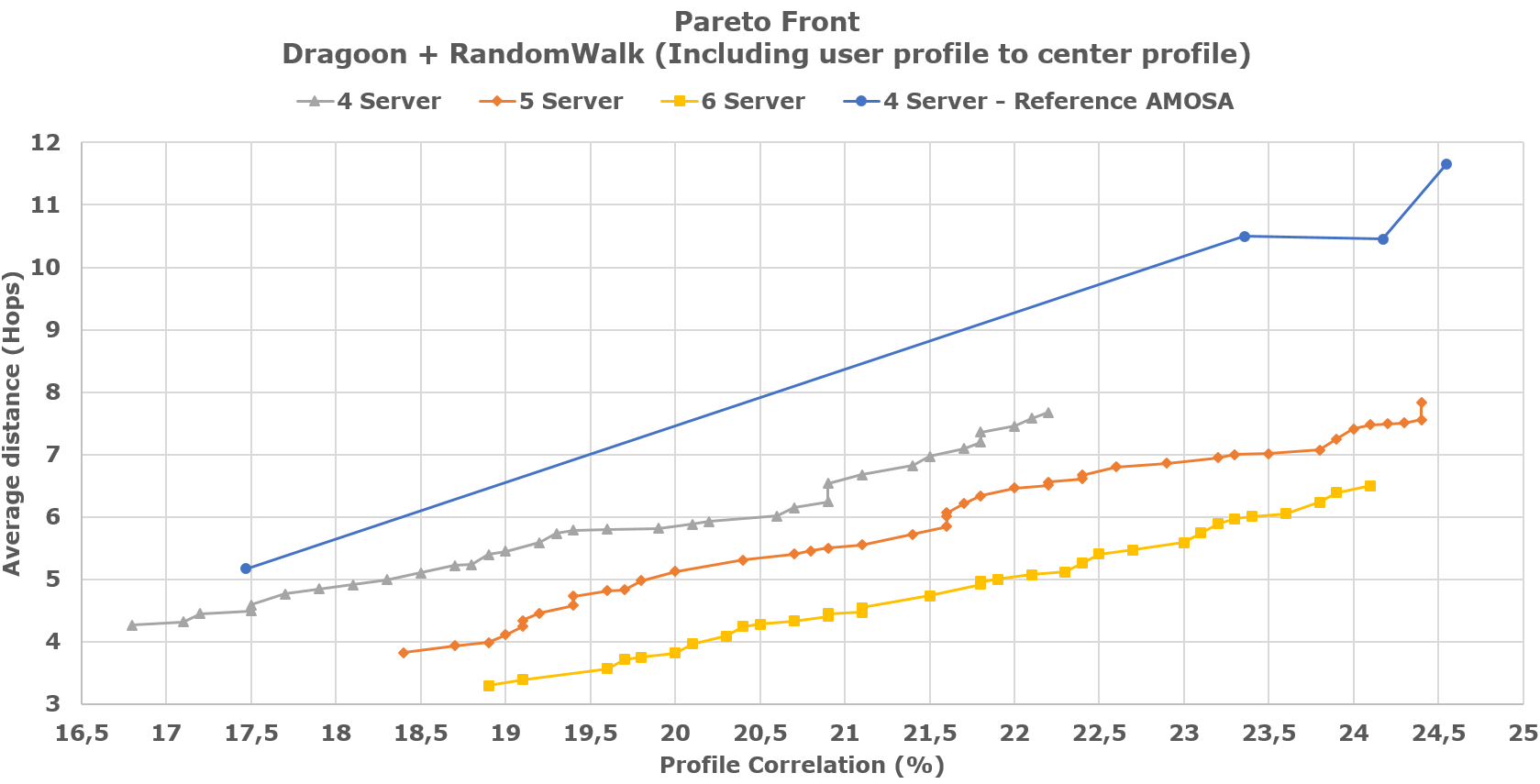}
\caption{Solutions at the local Pareto Front with algorithms Dragoon as initialization with optimized placement and Random Walk for iterative improvement towards profile correlation.}
\label{fig:exp4}
}\end{figure}

\textbf{Experiment 4:}
In line with our initial intention, we calculate a local Pareto Front with our approach for average network distance and profile correlation from a user to its primary server. The Figure \ref{fig:exp4} shows the effect of varying server numbers and their impact on the caching behavior as well as the network load.

The AMOSA algorithm calculated only 4 solutions at the Pareto Front. In contrast, our approach reached more fine grained solutions. The comparison for 4 servers shows that our combined optimization algorithms calculate a better Pareto Front.
With increasing amount of server, the average length of the network path is reduced. The ``optimal'' balance between distance and profile correlation depends on the goals of the CDN provider and ISP. It depends on agreements between the partners and respective calculations can be made for every point of the Pareto Front. The Pareto front provides a set of candidate solutions for the negotiations between CDN provider and ISP.
\section{CONCLUSION AND OUTLOOK}
In this paper, we propose a model to simulate and optimize a CDN according to multiple objectives. We model different user behaviors and take advantage of their profiles during optimization. Our described model considers multiple challenges in the storage chain, which are interacting with each other. We propose a novel approach to improve the entire processes. This enables an overall optimization to find combined solutions and highlights the cooperation between a CDN provider and ISP. With this model, we are able to simulate and analyze overall system processes and behaviors under specific constrains. We provide a tool to support the management in its decisions process with regard to the necessary amount of servers, the placement of these servers and the recommended cache size. Additionally, we improve the assignment of users to servers with fitting content. We simulated different system behaviors and visualized special aspects during runtime. Our analyses show that only a few servers are necessary in an optimized system to reach a high service level. In the future, we plan to include further organization strategies for the servers to balance the content and load.


\appendix

\bibliographystyle{wsc}
\bibliography{literature, demobib}

\section*{AUTHOR BIOGRAPHIES}
\noindent {\bf PETER HILLMANN} is a Ph.D. student at the Universit\"at der Bundeswehr M\"unchen (UniBwM), Germany. He holds a M.Sc. in Information-System-Technology from Dresden University of Technology since 2011. His areas of research are network and system security with focus on cryptography as well as operational modeling and optimization. His email address is \email{peter.hillmann@unibw.de}\\

\noindent {\bf TOBIAS UHLIG} is a postdoctoral researcher at the Universit\"at der Bundeswehr M\"unchen, Germany. He holds a M.Sc. degree in Computer Science from Dresden University of Technology and a Ph.D. degree in Computer Science from the Universit\"at der Bundeswehr M\"unchen. His research interests include operational modeling, natural computing and simulation-based optimization. He is a member of the ASIM and the IEEE RAS Technical Committee on Semiconductor Manufacturing Automation. His email address is \email{tobias.uhlig@unibw.de}\\

\noindent {\bf GABI DREO RODOSEK} is Professor of Communication System and Network Security at the Universit\"at der Bundeswehr M\"unchen, Germany. She holds a M.Sc. from the University of Maribor and a Ph.D. from the Ludwig-Maximilians University Munich. She is spokesperson of the Research Center Cyber Defence (CODE), which combines skills and activities of various institutes at the university, external organizations and the IT security industry (for instance Cassidian, IABG or Giesecke \& Devrient). Her email address is \email{gabi.dreo@unibw.de}.\\

\noindent {\bf OLIVER ROSE} holds the Chair for Modeling and Simulation at the Department of Computer Science of the Universit\"at der Bundeswehr M\"unchen, Germany. He received a M.S. degree in applied mathematics (1992) and a Ph.D. degree in computer science (1997) from W\"urzburg University, Germany. His research focuses on the operational modeling, analysis and material flow control of complex manufacturing facilities, in particular, semiconductor factories and assembly systems. He is a member of INFORMS Simulation Society, ASIM (German Simulation Society), and GI (German Computer Science Society). In 2012, he served as General Chair of the WSC. Currently, he is member of the board of the ASIM and the ASIM representative at the Board of Directors of the WSC. His email address is \email{oliver.rose@unibw.de}.\\

\end{document}